\documentclass[%
preprint,
superscriptaddress,
 amsmath,amssymb,
 aps,
]{revtex4-2}

\usepackage{graphicx}% Include figure files
\usepackage{dcolumn}% Align table columns on decimal point
\usepackage{bm}% bold math
\usepackage{physics}
\usepackage{xcolor}
\usepackage{soul}
\usepackage{braket}
\usepackage[]{changes}
\usepackage[colorlinks=true,
            linkcolor=red,
            urlcolor=red,
            citecolor=red]{hyperref}% add hypertext capabilities
\begin{document}

\preprint{APS/123-QED}

\title{Floquet quadrupole photonic crystals protected by space-time symmetry}

\author{Jicheng Jin}
\author{Li He}
\author{Jian Lu}
\author{Eugene J. Mele}
\author{Bo Zhen}
\email{bozhen@sas.upenn.edu}
\affiliation{%
 Department of Physics and Astronomy, University of Pennsylvania, Philadelphia, Pennsylvania, 19104, USA\\
 %This line break forced with \textbackslash\textbackslash
}%

\date{\today}% It is always \today, today,
             %  but any date may be explicitly specified

\begin{abstract}

High-order topological phases, such as those with nontrivial quadrupole moments\cite{benalcazar_electric_2017,benalcazar_quantized_2017}, protect edge states that are themselves topological insulators in lower dimensions. So far, most quadrupole phases of light are explored in linear optical systems, which are protected by spatial symmetries \cite{he_quadrupole_2020} or synthetic symmetries \cite{benalcazar_electric_2017,benalcazar_quantized_2017,dutt_higherorder_2020,mittal_photonic_2019,peterson_fractional_2020,peterson_quantized_2018}. Here we present Floquet quadrupole phases in driven nonlinear photonic crystals (PhCs) that are protected by space-time screw symmetries\cite{xu_spacetime_2018}. We start by illustrating space-time symmetries by tracking the trajectory of instantaneous optical axes of the driven media. 
Our Floquet quadrupole phase is then confirmed in two independent ways: symmetry indices at high-symmetry momentum points and calculations of the nested Wannier bands. Our work presents a general framework to analyze symmetries in driven optical materials and paves the way to further exploring symmetry-protected topological phases in Floquet systems and their optoelectronic applications.

\end{abstract}

\maketitle

%\tableofcontents

\section{introduction}
Symmetry plays an important role in topological phases~\cite{chiu_classification_2016,schnyder_classification_2008,kitaev_periodic_2009,ryu_topological_2010,lu_topological_2014,ozawa_topological_2019}. 
Examples include topological insulators that are protected by time-reversal symmetry~\cite{kane_quantum_2005,fu_topological_2007},
Chern insulators that require breaking time-reversal symmetry \cite{haldane_possible_2008a,wang_reflectionfree_2008,wang_observation_2009,skirlo_experimental_2015}, 
and topological crystalline insulators~\cite{fu_topological_2011} that are protected by spatial symmetries such as rotation and reflection. 
One important class of topological crystalline insulators is high-order topological insulators \cite{benalcazar_electric_2017,benalcazar_quantization_2019,peterson_fractional_2020,peterson_quantized_2018,liu_bulk_2021,zhang_secondorder_2019,zhang_symmetryprotected_2020,imhof_topolectricalcircuit_2018,serra-garcia_observation_2018}, where the interesting physical consequences appear in spaces two or more dimensions lower than the bulk. 
For example, quadrupole topological insulators in two dimensions, characterized by their quantized and non-trivial second-order moments, protect zero-dimensional corner states with fractional occupations.
So far, most studied quadrupole phases of light 
are protected by synthetic symmetries in the lattice model - such as the notion of $\pi$-fluxes~\cite{benalcazar_electric_2017,mittal_photonic_2019,peterson_quantized_2018} - or spatial symmetries such as the four-fold rotation~\cite{he_quadrupole_2020}. 
Moving beyond linear optics, a different class of Floquet topological phases can be found in nonlinear materials driven by time-varying fields. 
While some examples of Floquet topological phases have been explored \cite{dutt_higherorder_2020,rechtsman_photonic_2013,he_floquet_2019,fang_anomalous_2019}, 
detailed symmetry analysis, in both space and time, and the general recipe to achieve symmetry-protected topological phases in driven nonlinear optical systems remain largely unexplored. 
Here we present Floquet quadrupole phases in driven nonlinear PhCs, where the quadruple moments are quantized and protected by a space-time screw symmetry, involving both rotation in space and translation in time.  
We define and analyze this space-time symmetry in a specific example of driven GaAs before presenting the detailed design of a Floquet quadrupole PhC. 
The nontrivial quadrupole moment is then confirmed with the numerical calculations of the nested Wilson loops. 
Finally, we demonstrate key features associated with quadrupole phases, including fractional corner occupations and filling anomalies\cite{benalcazar_quantization_2019,he_quadrupole_2020}. 
%%%%%%%%%%    Figure 1     %%%%%%%%%%
\begin{figure}[h]
\includegraphics{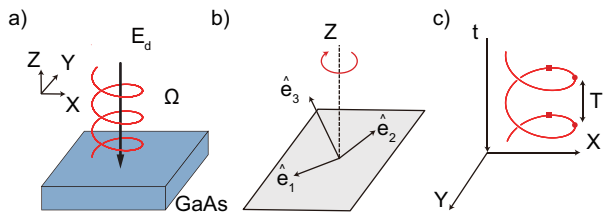} 
\caption{\label{fig1} 
{\bf Space-time screw symmetry in a driven nonlinear medium.}
a) Schematic of a uniform slab of GaAs driven by a circularly-polarized field ${\bf E}_{d}$ incident from the normal direction. 
b) 
All three instantaneous optical axes of the driven medium, $\hat{e}_{1-3}$, spin around the $z$ axis. 
c) Screw symmetry can be found in each of the optical axes, defined through a combination of rotation in space ($x$,$y$) and translation in time ($t$).}
\end{figure}

\section{Results}
\subsection{Space-time screw symmetry in driven materials}
We start by defining the space-time screw symmetry in a driven nonlinear medium. 
As shown in Fig.~\ref{fig1}a, a circularly-polarized field $\textbf{E}_d = E_d \cos(\Omega t) \hat x + E_d \sin(\Omega t) \hat y$ periodically drives a uniform slab of GaAs. 
This driving field couples to the second-order nonlinear susceptibility of GaAs, $\chi^{(2)}_{xyz}$ and its permutations, and gives rise to a time-dependent permittivity:  
\begin{equation}
\Bar{\Bar{\varepsilon}} (t) =\begin{pmatrix}
\varepsilon  & 0 & \alpha \sin( \Omega t)\\
0 & \varepsilon  & \alpha\cos( \Omega t)\\
\alpha\sin( {\Omega}t) & \alpha\cos( \Omega t) & \varepsilon 
\end{pmatrix}.
\end{equation}
Here, $\varepsilon$ is the linear permittivity of GaAs that is isotropic and $\alpha = \chi^{(2)}_{xyz} E_d$ is the nonlinear perturbation. 
Higher-order perturbations are ignored here under the assumption of a weak driving field.
To determine the symmetry of this driven medium, we analyze the temporal evolution of its three optical axes: 
\begin{align}
    \hat{e}_1 &=  \cos(\Omega t)\hat{x} - \sin(\Omega t)\hat{y} \notag \\
    \hat{e}_2 &=  \frac{1}{\sqrt{2}}( \sin(\Omega t)\hat{x} + \cos(\Omega t)\hat{y} - \hat{z}) \notag \\
    \hat{e}_3 &=  \frac{1}{\sqrt{2}}(\sin(\Omega t)\hat{x} + \cos(\Omega t)\hat{y} + \hat{z}) \
\end{align}
As shown in Fig.~\ref{fig1}b, all three optical axes spin around the $z$ axis at the driving frequency $\Omega$.
We can further trace out the trajectory of the optical axes in both space ($x,y$) and time ($t$). 
An example for $\hat{e}_1$ is shown in Fig.~\ref{fig1}c, which evolves along a helix. This provides the foundation for our symmetry analysis below. 

First, we note that the driving field breaks the continuous rotation symmetry of the isotropic linear permittivity of GaAs; 
namely, the helix in Fig.~\ref{fig1}c does not return to itself if it is rotated by some general angle in the $xy$ plane  (e.g. $90$ degrees). Instead, it has a symmetry involving a compound operation with a rotation in space and a shift in time. For example, one can first rotate the helix by $90$ degrees along the counterclockwise direction in the $xy$ plane ($C_4$) and then translate it by $T/4$ in time ($\hat{T}_{T/4}$). Here $T = 2\pi/\Omega$ is the periodicity of the driving field. 
For convenience, we denote this space-time screw operation as:
\begin{equation}
    \tilde{S}_4 = \hat{O}_{C_4} \times \hat{T}_{T/4}.
\end{equation}
Naturally, if one repeats this $\tilde{S}_4$ operation four times, the whole system evolves in time by a full periodicity $T$ and, thus, remains unchanged. 
Under this requirement of $(\tilde{S}_{4})^4 = 1$, the four allowed $\tilde{S}_4$ symmetry indices are $\pm 1$ and $\pm i$. 
The symmetry of this driven medium can also be derived by checking the commutation rules between various symmetry operations and the time-dependent nonlinear permittivity, reaching the same conclusions. 
See Section I of the Supplementary Information for more details. 
We stress that, as the symmetry analysis is on the effective permittivity, it is not uniquely defined by the driving field; instead, it also depends on the exact form of optical nonlinearity provided by the material. 
For example, under the exact same driving field, a $z$-cut LiNbO$_3$ slab has a different space-time screw symmetry of $\tilde{U}_4 = \hat{O}_{C_4} \hat{T}_{-T/4}$ with a different time-translation term. 
We further note that while all monochromatic driving fields have space-time symmetry, it is not always true in driven materials. 
For example, an $x$-cut LiNbO$_3$ driven by a $z$-polarized field will have not a space-time symmetry but a purely spatial symmetry of $C_2^x$ . 
See Section II of the Supplementary Information for the symmetry analysis of these mentioned scenarios.  

\begin{figure}[h]
\includegraphics{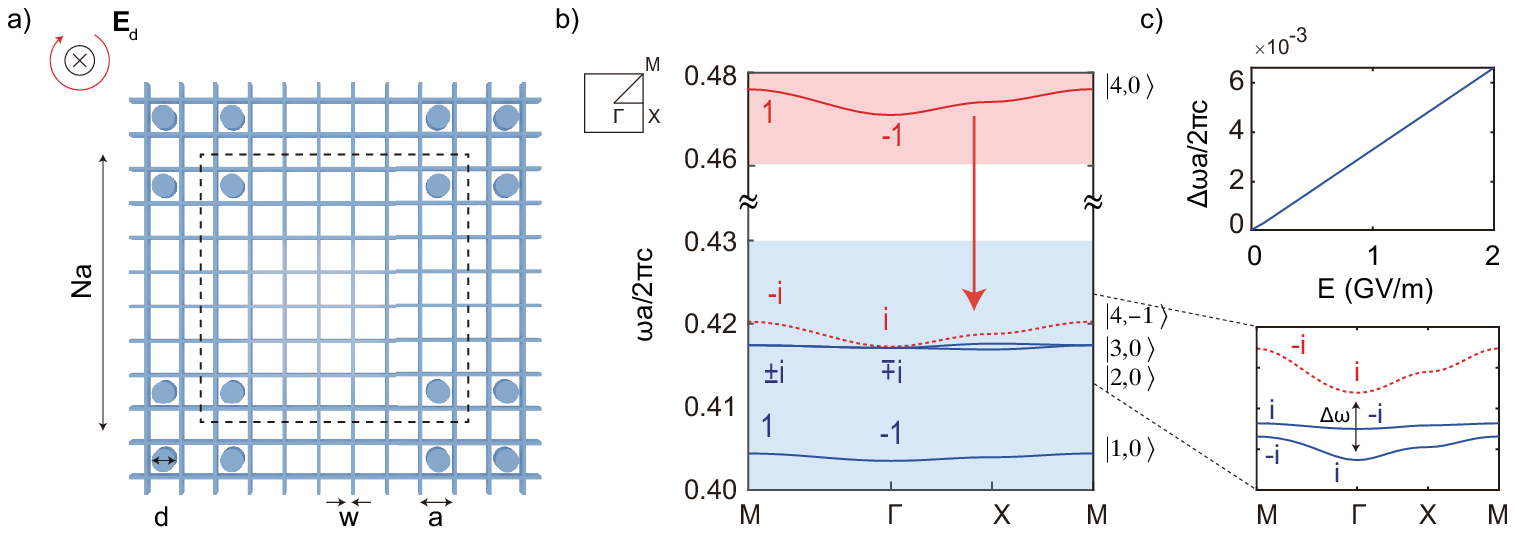} 
\caption{\label{fig2} 
{\bf Space-time symmetry indices of photonic bands dressed by optical nonlinearity.}
a) 
Schematic drawing of a square GaAs PhC unit cell, consisting of veins of width $w$ and spacing $a$ as well as 4 disks of diameter $d$. 
b) 
The PhC band structure, including both TE (blue lines) and TM bands (red). 
The $\tilde{S}_4$ symmetry indices are labeled at the high-symmetry momentum points  ($\Gamma$ and $M$) of each band. 
If driven by an external field, Floquet basis sharing the same $\tilde {S}_4$ index, such as $\ket{4,-1}$ and $\ket{2,0}$ at $\Gamma$,  will couple to each other, which leads to an energy splitting of $\Delta \omega$ and opens a new gap in the Floquet spectrum.  
c) 
The energy splitting $\Delta \omega$ increases linearly with the driving field strength $E$. 
}

\end{figure}

\subsection{Floquet quadrupole photonic crystal design}
Next, we present a concrete example of Floquet quadrupole photonic crystal (PhC) that is protected by this space-time symmetry of $\tilde{S}_4$. 
As shown in Fig.~\ref{fig2}a, the 2D PhC consists of veins and disks made from GaAs in the air, and one of the unit cells is defined by the black dashed lines. 
Veins of width $w$ = 50 nm form a square lattice of periodicity $a$ = 500 nm. 
Four disks of diameter $d$ = 348 nm are arranged in a $C_4$ symmetric way in each unit cell. 
The calculated PhC band structure, eigen-frequencies $\omega$ as functions of momentum $k$, is shown in Fig.~2b, where TE modes ($E_x,E_y,H_z$) and TM modes ($H_x,H_y,E_z$) are colored in blue and red, respectively. 
By engineering the location and size of the disks, four of the bands %($1-4$) 
are well isolated from the rest, each residing inside a TE or TM band gap that is shaded in blue or red. 
More details of the numerical simulation can be found in Methods.  

Similar to our previous analysis, a circularly polarized driving field ${\bf E}_d$ breaks the spatial rotation symmetry $C_4$ in this PhC. Instead, the PhC now has
a space-time screw symmetry $\tilde{S}_4$, which quantizes bulk dipole and quadrupole moments. 
See Section III in the Supplementary Information for detailed derivations. 
The bulk quadrupole moment $q_{xy}$ of two isolated bands can be evaluated using the $\tilde S_4$ symmetry indices at their high-symmetry momentum points as: 
\begin{equation}
\label{eq1}
    e^{i2\pi q_{xy}} = \tilde{S}^{+}_4(\Gamma) \tilde{S}^{+*}_4(M) = \tilde{S}^{-}_4(\Gamma) \tilde{S}^{-*}_4(M). 
\end{equation}
Here $\tilde{S}_4^{\pm} $ refers to the $\tilde{S}_4$ eigenvalue of the mode with $(\tilde{S}_4)^2$ eigenvalue of $\pm 1$, so naturally, $\tilde{S}_4^{+} = \pm 1$ and $\tilde{S}_4^{-} = \pm i$. 

To find the Floquet eigenstates of this driven PhC and their relevant $\tilde{S}_4$ indices, we solve the Floquet eigenvalue problem of Maxwell's equations, following our previous theoretical framework \cite{he_floquet_2019}. 
In short, we expand the Floquet eigenstates using the Floquet basis as $\Phi(t) = e^{-i\lambda t} \sum _ {jm} c_{jm}\ket {j,m}$ and then compute the Floquet eigenvalues $\lambda$ and coefficients $c_{jm}$. 
Floquet basis states $\ket{j,m} = \ket{j} e^{im\Omega t}$ are essentially copies of the static PhC eigenstate $\ket{j}$, but shifted in frequency by $m \Omega$. 
One example is the Floquet basis $\ket{4, -1}$ shown in Fig.~\ref{fig2}b, which is shifted down by $\Omega$ from $\ket{4, 0}$. 

To understand the $\tilde{S}_4$ indices of the Floquet eigenstates, we start by comparing the symmetries under $C_4$ and the compound operation $\tilde{S}_4$.
\begin{eqnarray}
      \hat{O}_{\tilde{S}_4}\ket{j,m} &=& (\zeta_j \times i^m )\ket{j,m},\\ 
      \hat{O}_{\tilde{S}_4}\ket{j,0} &=& \zeta_j \ket{j,0}.\
\end{eqnarray}

Namely, the $\tilde{S}_4$ index depends on the band information $j$ and the Floquet order $m$.
For example, for $m=0$, the $\tilde{S}_4$ index reduces to $C_4$ index $\zeta_j$ of $\ket{j}$; 
for $m=\pm 1$, the $\tilde{S}_4$ index is changed by $\pm i$. Our detailed derivations can be found in Section III of the Supplementary Information. Naturally, the $\tilde{S}_4$ index of a Floquet eigenstate is the same as that of all of its constituting Floquet basis. 

We now apply this symmetry analysis to our specific setup and compute the quadrupole moment.  
When the Floquet basis $\ket{4,0}$ shifts down in frequency to $\ket{4,-1}$, its $\tilde{S}_4$ index at $\Gamma$ changes from $-1$ to $+i$, which is now the same as the $\tilde{S}_4$ index of $\ket{2,0}$. 
Naturally, the two Floquet basis, $\ket{4,-1}$ and $\ket{2,0}$, will couple to each other under a driving field, resulting in an energy splitting $\Delta \omega$ between them. 
This energy splitting increases linearly with the driving field strength (Fig.~\ref{fig2}c), lifts the degeneracy between static states $\ket{2,0}$ and $\ket{3,0}$, and opens a new (Floquet) energy gap.
Using Eq.~\ref{eq1}, the quadrupole moment of the two bands below the Floquet gap, $\ket{1,0}$ and $\ket{2,0}$, can be evaluated as: $e^{i 2\pi q_{xy}} = -1$ and $q_{xy} = 1/2$; namely, we have now achieved a Floquet quadrupole phase.    

%%%%%%%%%%%%%%%%%%%%%%%%%%%%%%%%%%%%%

%%%%%%%%%%    Figure 3     %%%%%%%%%%
\begin{figure}[ht]
\includegraphics{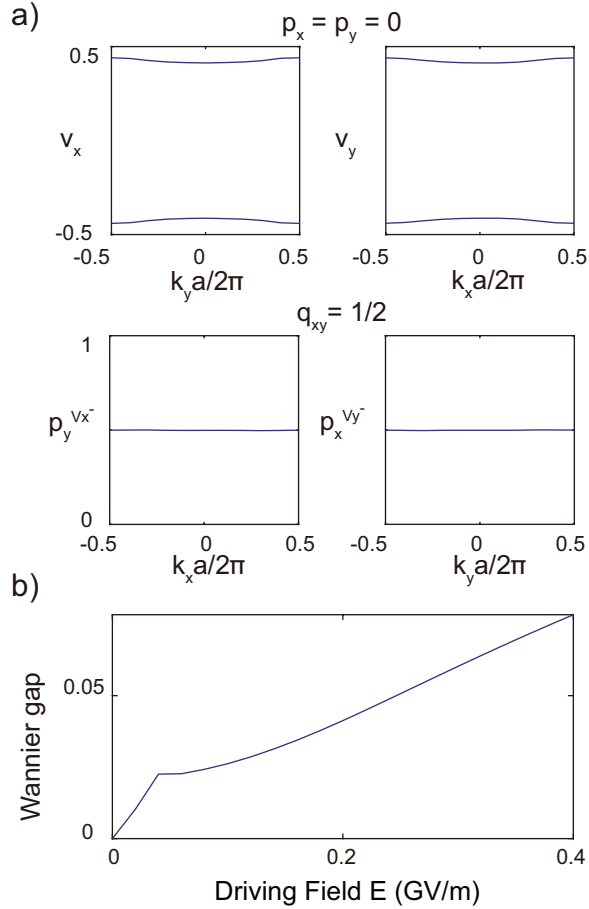}
\caption{\label{fig3} 
{\bf Confirmation of quadrupole phases through Wannier band calculations.}
a) Calculated Wannier bands and nested Wannier bands for the first two Floquet bands. The results confirm the vanishing dipole moments $p_{x}=p_{y}=0$ and the nontrivial quadrupole moment $q_{xy} = 1/2$. 
b) The gap between the two Wannier bands is opened by the external driving field.  
}
\end{figure}
%%%%%%%%%%%%%%%%%%%%%%%%%%%%%%%%%%%%%

Next, we confirm the Floquet quadrupole phase through direct calculations of
the nested Wannier bands. 
To this end, we first compute the Wannier bands $v_{x,y}$ of the two Floquet bands of interest, $\ket{1,0}$ and $\ket{2,0}$. 
Our results, shown in the upper panel of Fig.~\ref{fig3}a, confirm that we have vanishing dipole moments in both directions: $p_x = p_y = 0$. 
Besides an energy gap (Fig.~\ref{fig2}c), the driving field also opens a Wannier gap between the two Wannier bands (Fig.~\ref{fig3}b), which are gapless without a driving field. 
This Wannier gap allows one to separate the Wannier bands into two sectors, $v^{\pm}$, and compute the nested Wannier bands. 
Our results, shown in the lower panel of Fig.~\ref{fig3}a, confirm that our driven PhC indeed has a nontrivial quadrupole moment of $q_{xy} = 2 p_x^{v_y^-} p_y^{v_x^{-}} = 1/2$. 

%%%%%%%%%%    Figure 4     %%%%%%%%%%
\begin{figure}[ht]
\includegraphics{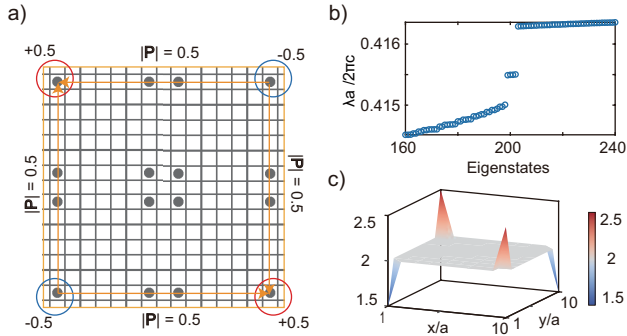}
\caption{\label{fig4} 
{\bf Physical consequences of corner states and filling anomaly.}
a) Schematic of a $10 \times 10$ super-cell of Floquet quadrupole PhCs. 
b) The eigenvalue spectrum confirms the existence of corner states and the filling anomaly in our system.  
c)  Accumulative time-averaged energy density profile of the first 200 eigenstates, showing fractional occupations ($2 \pm 0.5$) at the four corners. 
}
\end{figure}
%%%%%%%%%%%%%%%%%%%%%%%%%%%%%%%%%%%%%

\subsection{Physical consequences of quadrupole phases}
Finally, we present the physical consequences of Floquet quadrupole PhCs in the contexts of corner states and filling anomalies. 
We start by computing the eigenstates in a $N \times N$ super-cell of Floquet quadrupole PhCs surrounded by perfect electric conductors (PECs). 
Our specific setup with $N=10$ is shown in Fig.~\ref{fig4}a. 
There is a thin gap between the super-cell and the PECs.
The eigenstates are labeled in the order of their Floquet eigenvalues.
Similar to other quadrupole phases (refs), we also observe 4 degenerate states in energy gap (states 199 - 202 in Fig.~\ref{fig4}b), which are localized at the 4 super-cell corners.  
Due to the lack of chiral symmetry expanded around a nonzero frequency in Maxwell's equations, these corner states are not pinned to the center of the energy gap; instead, they can shift up or down in frequency or even merge into the bulk continuum. 
A filling anomaly is also confirmed in our system by noting the incompatibility between the number of eigenstates below the Floquet gap ($2N^2 -2 = 198$) and the number of unit cells in the super-cell ($N^2 = 100$). 
Our quadrupole phase is further confirmed by the fractional occupations at the corners, which is an integral of the mode density over the occupied bands, as shown in Fig.~\ref{fig4}c. 
In these calculations, the disk diameter $d$ is tuned to place the corner-state frequency in the middle of the Floquet gap. Details of the calculation are presented in Section IV of the Supplementary Information.

\section{Discussion}
In summary, we present Floquet quadrupole phases that are protected by the space-time screw symmetry in a driven nonlinear PhC. 
The parameters used in our calculations are practical, and the proposed system can be readily studied 
in nonlinear optical experiments. 
Furthermore, while our example focuses on GaAs, the space-time symmetry analysis applies to the vast range of nonlinear materials, opening the door to further explorations into new phases and consequences in nonlinear topological photonics.
Finally, our general formalism of understanding Floquet topological phases in driven systems can extend beyond photonics into other nonlinear wave systems, including phononics, piezoelectrics, piezomagnetics, and polaritonics. 
\section{Acknowledgement}
This work was partly supported by the National Science Foundation through the University of Pennsylvania Materials Research Science and Engineering Center DMR-1720530, the US Office of Naval Research (ONR) Multidisciplinary University Research Initiative (MURI) grant N00014-20-1-2325 on Robust Photonic Materials with High-Order Topological Protection, and the Air Force Office of Scientific Research under award number FA9550-18-1-0133.  J.L. was also supported by the Army Research Office under award contract W911NF-19-1-0087. Work by E.M. was supported by the Department of Energy, Office of Basic Energy Sciences under grant DE FG02 84ER45118.

\bibliography{reference}% Produces the bibliography via BibTeX.

\end{document}